# Observation of Magnetic Skyrmion Bubbles in a van der Waals ferromagnet Fe$_3$GeTe$_2$

Bei Ding[†,⊥,*], Zefang Li[†,⊥,*], Guizhou Xu[‡,*], Hang Li[†,⊥], Zhipeng Hou[§], Enke Liu[†], Xuekui Xi[†], Feng Xu[‡], Yuan Yao[†], and Wenhong Wang[†,#]

[†]Beijing National Laboratory for Condensed Matter Physics, Institute of Physics, Chinese Academy of Sciences, Beijing 100190, China

[‡] School of Materials Science and Engineering, Nanjing University of Science and Technology, Nanjing 210094, China

[§]South China Academy of Advanced Optoelectronics, South China Normal University, Guangzhou 510006, China

[⊥]University of Chinese Academy of Sciences, Beijing 100049, China

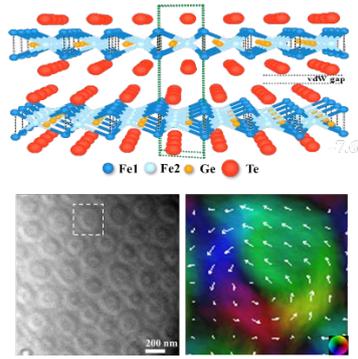

**ABSTRACT:** Two-dimensional (2D) van der Waals (vdW) magnetic materials have recently been introduced as a new horizon in materials science and enable the potential applications for next-generation spintronic devices. Here, in this communication, the observations of stable Bloch-type magnetic skyrmions in single crystals of 2D vdW $Fe_3GeTe_2$ (FGT) are reported by using *in-situ* Lorentz transmission electron microscopy (TEM). We find the ground-state magnetic stripe domains in FGT transform into skyrmion bubbles when an external magnetic field is applied perpendicularly to the (001) thin plate with temperatures below the Curie-temperature $T_C$. Most interestingly, a hexagonal lattice of skyrmion bubbles is obtained via field cooling manipulation with magnetic field applied along the [001] direction. Owing to their topological stability, the skyrmion bubble lattices are stable to large field-cooling tilted angles and further reproduced by utilizing the micromagnetic simulations. These observations directly demonstrate that the 2D vdW FGT possesses a rich variety of topological spin textures, being of a great promise candidate for future applications in the field of spintronics.

Two-dimensional (2D) van der Waals (vdW) materials are a family of quantum materials that have attracted great research attention in the past decade as they possess a diverse range of novel phenomena which are promising for technological applications. [1, 2] In particular, the recent discovery of magnetic 2D vdW materials, such as $Cr_2Si_2Te_6$/$Cr_2Ge_2Te_6$, [3-5] $CrI_3$/$CrBr_3$, [6, 7] and $Fe_3GeTe_2$ (FGT), [8, 9] not only offers exciting opportunities for exploring new physical properties, but also opens up a new way for developing spintronic devices by applying magnetism as a possible altering parameter. [10] Among these materials, FGT is only ferromagnetic metal, in which a long-range ferromagnetic order has been confirmed experimentally ranging from bulk crystals down to monolayers. [11-13] Remarkably, bulk crystalline FGT has the highest Curie temperature $T_C$ (~230 K) and the $T_C$ of layered FGT can be raised to room temperature via electrostatic gating [8, 14] or in patterned microstructures. [13] Following this discovery, many intriguing magnetic and transport properties, such as extremely large anomalous Hall effect, [15] Planar topological Hall effect, [16] Kondo lattice physics, [17] anisotropy magnetostriction effect, [18] and spin filtered tunneling effect, [19] have been observed experimentally in exfoliated FGT nanoflakes and its heterostructures.

Moreover, 2D vdW FGT exhibits a strong out-of-plane uniaxial magnetic anisotropy down to atomic-layer thicknesses, [8, 9, 14, 20] which is very critical for spintronic applications, typically, magnetic-tunneling-junctions and magnetic random-access-memory devices. On the other hand, in a magnetic material, the competition between the uniaxial magnetic anisotropy and magnetic dipole-dipole interaction, can emerge and lead to a diversity of topological spin configuration that are defined with their unique topological number. [21] For example, magnetic skyrmionic bubbles with various topological spin textures have been experimentally discovered in a range of centrosymmetric magnets, such as perovskite manganites, [21, 22] hexagonal MnNiGa alloys, [23-25] and layered kagome $Fe_3Sn_2$ magnets. [26, 27] More importantly, recent theoretical and experimental works have discussed the emergence of topological spin textures including magnetic skyrmions in 2D vdW materials and their heterostructures for future spintronic applications. [28, 29] These findings inspire us to investigate the magnetic domain structures and magnetization dynamics of 2D vdW FGT crystals using Lorentz transmission electron microscopy (TEM).

Here, in this work, we report on a Bloch-type magnetic skyrmion bubble which can indeed be observed in the single crystal sample of 2D vdW FGT. Owing to the competition between uniaxial magnetic anisotropy and magnetic dipole-dipole interactions, at temperatures below $T_C$, we found that the magnetic stripe domains in FGT thin plates turn into magnetic skyrmion bubbles with magnetic field applied perpendicularly to the (001) plane. Moreover, a high-density hexagonally-packed lattice of skyrmion bubbles emerges by a simple field-cooling process and keeps stable after reducing the magnetic field to zero. Owing to their topological stability, the skyrmion bubble lattices are stable to large field-cooling tilted angles and further reproduced by utilizing the micromagnetic simulations. These observations directly demonstrate that the 2D vdW FGT possesses a rich variety of topological spin textures, being of a great promise candidate for future applications in the field of spintronics.

Single crystal $Fe_3GeTe_2$ were synthesized by using the self-flux technique with a mixture of pure elements Ge (99.9999%), Fe (99.99%) and Te (99.995%) [see Methods for details]. As schematically shown in **Figure 1a, b**, 2D vdW FGT belongs to space group *P63/mmc* with two Te layers separated by a layered $Fe_3Ge$ hetero-metallic slabs. There are two inequivalent Fe(1) and Fe (2) atoms in the $Fe_3Ge$ slabs, forming two separated triangular lattice Fe(1)-Fe(1) layer and a Fe(2)-Ge hexagonal atomic ring layer. [11] **Figure 1c** shows the typical X-ray diffraction (XRD) pattern of a bulk FGT single crystal which agrees with the recent work.[30] One can notice that the XRD pattern represents only the (0 0 2n) Bragg peaks (n=1,2,3,4,5,6), indicating that the studied surface is the *ab* plane of FGT crystal. The crystal structure of FGT crystals was further examined by atomic resolution scanning transmission electron microscopy (STEM), as shown in **Figure 1d, e**. The bi-layer Te atoms are clearly observed in STEM HAADF image along the [100] axis as shown in **Figure 1d**, confirming the layered structure and high quality of our samples. In [001] projection imaging (**Figure 1e**), we clearly see the arrangement of the hexagonal ring of Fe(2)-Ge layer and the triangular structure of Fe(1) atoms. The thickness and composition of the FGT thin plate was determined by electron energy loss spectrum (EELS) and scanning energy dispersive x-ray (EDX) spectroscopy, demonstrating the uniform element distribution of Ge, Fe, and Te across the surface (see **Figure S1 and S2, SI**).

In **Figure 1f,** the zero field cooling (ZFC) and field cooling (FC) M(T) curves are measured with external field H = 100 Oe applied both parallel to the *c* direction and in the *ab* plane. Clearly, a paramagnetic to ferromagnetic transition approximately observed at $T_C$ ~ 150 K with both directions, consistent with previous reported for the flux-grown samples. [31, 32] The magnetization of the two direction (H//ab and H//c) exhibit a distinctive difference which is caused by the magnetic anisotropic character of FGT crystals. Moreover, when H//c, the FC and ZFC plots occur a bifurcation below $T_C$ similar as other typical frustrated magnets [26, 27, 33]. The inset of **Figure 1f** show the magnetization curves measured at $T$ = 10 K with H//ab and H//c, respectively. It can be seen that the magnetic easy axis is along H//c revealing a strong magnetic anisotropy in FGT. The saturated magnetic moment ($M_S$) is 3.25 $\mu_B$ / f. u. consistent with the reported values. [11, 31, 32]

**Figure 2a-d** show the evolution of magnetic domain structures as a function of temperature observed by using *in-situ* Lorentz-TEM under zero magnetic field (see **Supplementary Movie 1, SI**). The corresponding view area is along [001] direction which is presented though selected-area electron diffraction (SAED) measurements (see the inset of **Figure 2a**). We noticed that, at temperatures below $T_C$ ~ 150 K, the magnetic stripe domains emerge with an average width of ~120 nm as shown in **Figure 2b**. This value is comparable to that in the MnNiGa [23] and $Fe_3Sn_2$,[26] while roughly two times larger than in the $La_{1-x}Sr_xMnO_3$ (x=0.175). [22] The characteristic bright and dark magnetic contrast appears in the sample indicating that the spin of the magnetic stripe domains stands upward and downward along the magnetic easy axis which are separated by Bloch domain walls. As the temperature decreased, the magnetic stripes became widen and distinct while the domain wall still remained the original state. One should note that, the critical temperatures $T_c$ of the Lorentz-TEM sample is consistent with that of the bulk samples.

To confirm the variation of magnetic stripe domains under an application of magnetic field, we then investigated the magnetic domain texture evolution under different magnetic fields at $T$ = 110 K, shown in **Figure 2e-h**. The magnetic field was applied along [001] direction by increasing the objective lens current. The corresponding dynamics of formation and disappearance of the skyrmion bubbles with increasing applied magnetic field were successfully recorded by using *in-situ* Lorentz TEM (**see Figure S3, SI**). Remarkably, the magnetic stripe domain structure gradually transformed into skyrmion bubbles is spotted with the magnetic field increasing from 0 to 920 Oe. At a lower magnetic field, the magnetic stripe domain parallels to external field expanding at expense with the antiparallel ones. **Figure 2e** displays an image of the transformation procedure under a magnetic field of 360 Oe. It is clearly that the stripe domains, fragmentary magnetic domains together with skyrmion bubbles existing in this state. With the increasing of magnetic field (**Figure 2f-h**)**,** the evolution from residual magnetic stripe domains to the dumbbell-shaped magnetic domains first gradually occurs before totally shrinking into the skyrmion bubbles. As the magnetic field increases further above 680 Oe, the dumbbell-shaped magnetic stripes are completely replaced by skyrmion bubbles. With further increasing magnetic field, the size of skyrmion bubbles decreases and eventually disappears in the ferromagnetic state (**Figure 2h**) (See **Figure S4 for detailed analysis, SI**). These observations clearly indicate that isolated skyrmion bubble can be obtained in the 2D vdW FGT magnet by applying the magnetic field perpendicular to the thin plate. The saturated magnetic field of the Lorentz-TEM sample coincides with that of the bulk sample measured by magnetization (see **Figure S5, SI**), which can arise from the fact that the magnetic anisotropy in FGT is high enough to overcome the increase of demagnetizing energy in the thin Lorentz-TEM sample. At $T = 10$ K, the uniaxial magnetic anisotropy constant $K_u$ of FGT single crystals is calculated to be as large as ≈$10^7$ erg cm$^{-3}$, consistent with the reported values. [10]

**Figure 3a-d** show the detailed evolution of magnetic domain structures as a function of temperature at a fixed magnetic field of 600 Oe, the specific field-cooling (FC) manipulation is same as our previous work.[34] Although the Lorentz-TEM image appears some scratches due to the iron milling which further results in a slightly distorting magnetization distribution, it will not affect the observation of the bubble lattices. The magnetic domain appears contrast at ~124 K (**Figure 3b**) and gradually forms a single skyrmion bubble as the temperature is about 116 K (**Figure 3c**). When we decrease the temperature to 93 K (below $T_C$ ~ 150 K) during the FC process, a hexagonal lattice of skyrmion bubbles is obtained as shown in **Figure 3d**. Most interestingly, the bubble lattices keep unchanged after turning off the field to zero (**Figure 3e**). The zero-field stabilized lattice of skyrmion bubbles is probably a metastable state, during the zero-field warming (ZFW) process, as shown in **Figure 3e-g**, causing a little bit of size decrease and then gradually changing into the stripe domains (see **Supplementary Movie 2, SI**). As approaching $T_C$, whereas the magnetic contrast become weaker at ~ 134 K (**Figure 3h**) due to the decreased amplitude of the magnetization and increased thermal disorder. These results are similar to the skyrmion bubble lattices reported in other centrosymmetric magnets, such as tetragonal

perovskite magnetites [22] and hexagonal MnNiGa alloys. [24] Thus, it turns out that the magnetic fields play a critical role during the FC process to generate the highest-density hexagonal lattice of skyrmion bubbles. On the other hand, we should point out that, after the FC manipulation, the zero-field stabilized bubble lattices are found for the entire temperature range down to 93 K. This result is similar to that MnNiGa alloy, [24] but intrinsically different from that of chiral magnets, where skyrmion phase exists only in the narrow area below $T_C$. [35-37]

In the centrosymmetric magnets, the competition between the perpendicular magnetic anisotropy and magnetic dipole-dipole interaction is the crucial point in the generation of skyrmion bubbles. As a result, the spin textures of skyrmion bubbles in these materials are highly sensitive to the direction of the applied magnetic field related to the easy-axis. [38] To examine the stability of the skyrmion bubbles observed in FGT magnet in detail, we have further preformed the FC Lorentz-TEM experiments with a fixed magnetic field of 600 Oe applied at various oblique angles, i.e., the FGT sample was tilted with $\alpha = \pm 20°$ along the x-axis and $\beta = -10°$ along the y-axis, respectively. **Figure 4a-d** show the magnetic configuration observed at zero-field after various FC manipulations mentioned above. Remarkably, the FC procedure at different tilted angles can generate skyrmion bubbles and arrange them into a hexagonal lattice. These results agree well with a theoretical simulation of the bubble lattices generation in a small oblique field is shown in **Figure S6, SI**.

In order to better clarify the spin configuration of the skyrmion bubbles observed in FGT magnet, in **Figure 4e and 4f**, we show the under- and over-focused Lorentz

TEM images of zero-field stabilized lattice of skyrmion bubbles, respectively. The sharp contrast variations from dark to bright ring in the under-focused image identified with the Bloch domain walls separating the spin-up and spin-down domains. While it represents a reversal image contrast in the over-focused image shown in **Figure 4f**. The in-plane magnetization distribution map based on transport of intensity equation (TIE) analysis for a selected skyrmion bubble is shown in **Figure 4g** (See **Figure S7** for the TIE analysis of skyrmion bubble lattice), where white arrows indicate the size and direction of the magnetic component at each point. Clearly, the topological spin texture of a skyrmion bubble rotated counterclockwise forming a vortex-like magnetic domain, which is consistent with the Bloch-type skyrmion observed in both non-centrosymmetric (such as FeGe [39] and MnSi [40]) and centrosymmetric (such as $Fe_3Sn_2$ [26] and Ba-Fe-Sc-Mn-O [21]) magnets. These experimental results agree well with our theoretical simulations of the skyrmion bubbles in FGT magnets, as shown in **Figure 4h**, where the enclosed circular shape magnetic structures are revealed, signifying the presence of Bloch-type skyrmions.

In conclusion, we have demonstrated the formation of Bolch-type skyrmions in the 2D vdW FGT crystals. Owing to their topological stability, the densely hexagonal lattices of skyrmion bubble are stable to large tilted angles at zero field after subjecting a field-cooling manipulation. The stable skyrmion bubbles lattices that we observed are important for the exploitation of 2D vdW skyrmion-based devices and are likely to have a strong effect on magnetic and magnetotransport properties in these 2D vdW crystals. It should be point out that, during the submission of this manuscript, the Bolch-type skyrmion bubbles are also observed in another 2D vdW magnet $Cr_2Ge_2Te_6$.[41] Moreover, the observation of the Neel-type skyrmions in a thinner FGT film and its heterostructure are recently reported on the arXiv by T. E. Park *et al*. , Y. Wu *et al*. and H. Wang *et al*. in Refs. 42, 43 and 44, respectively. We thus anticipate that the tunable nature of 2D vdW magnets will enable the formation of a wide range of topological spin textures, by varying the thickness, the spin-orbit coupling, the interfacial symmetry of heterostructure, and thereby the magnetocrysalline anisotropy and the magnetization.

■ASSOCIATED CONTENT

**Supporting Information**

Additional experimental details including single crystals growth, structure and magnetic measurement, Curie temperature determination, magnetic-stripe domains and formation of magnetic skyrmion bubbles, micromagnetic simulation results in single crystals of $Fe_3GeTe_2$.

■AUTHOR INFORMATION

**Corresponding Author**

*E-mail: wenhong.wang@iphy.ac.cn

**Author Contributions**

B. Ding, Z. Li and G. Xu contributed equally to this work.

**Notes**

The authors declare no competing financial interests.

■ACKNOWLEDGMENT

This work is supported by the National Key R&D Program of China (2017YFA0303202 and 2017YFA0206303), the National Natural Science Foundation of China (11604148 and 11874410), and Beijing Natural Science Foundation (Grant No. Z190009)

**Figures and captions**

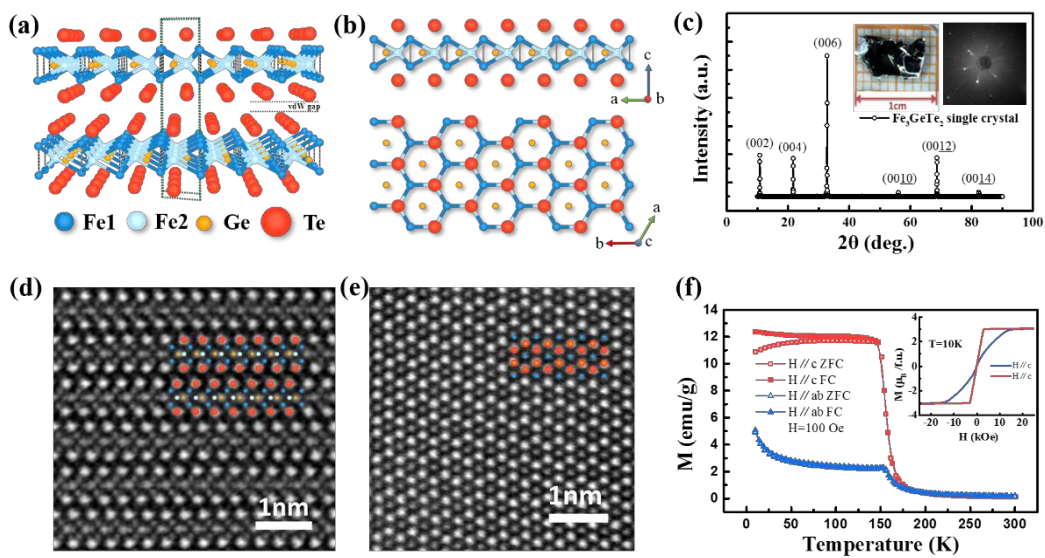

**Figure 1**. Structure and magnetic properties of a van der Waals (vdW) Fe$_3$GeTe$_2$ (FGT) single crystal. **a)** Schematic the structure of a FGT bi-layer with an interlayer vdW gap; **b)** the crystal structures of monolayered FGT viewed from *ac* and *ab* planes, respectively; **c)** XRD pattern for the as-grown slice of Fe$_3$GeTe$_2$ single crystal. The optical photograph and Laue image in the inset shows the typical size and Laue diffraction pattern, indicating the [001] orientation of the single crystal; **d, e)** high-resolution STEM HAADF images along the [010] and [001] directions, respectively. The insets show the arrangement of the stacking structure of Te atoms and the hexagonal ring of Fe$_{II}$-Ge layer. The scale bar is 1 nm; **f)** temperature dependence of the ZFC and FC magnetization measured at H = 100 Oe for H//*ab* and H//*c*. Inset shows the anisotropic M-H curve at T = 10 K.

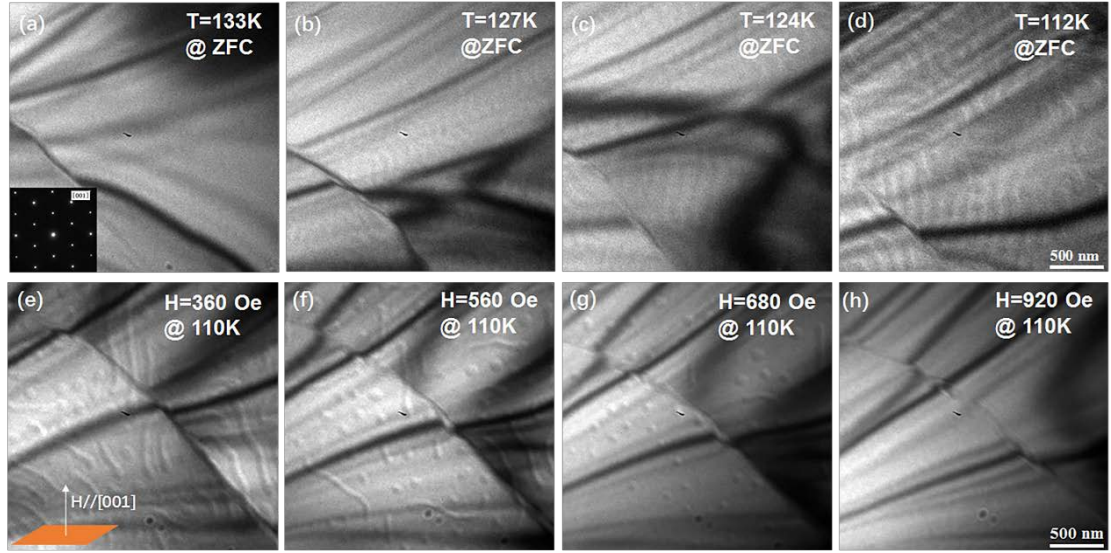

**Figure 2**. The representative images of the domain structures of a vdW FGT thin-plate taken by Lorentz-TEM with an electron beam perpendicular to the *ab*-plane. **a-d)** The under-focused Lorentz-TEM images of FGT when the sample temperature was lowered from 300 K to 100 K in a zero-field-cooling (ZFC) process. The inset of (a) shows the corresponding selected-area electron diffraction (SAED) pattern. At temperatures below 130 K, a spontaneous ground state of stripe domain appears and becomes more clear at a lower temperature of 112 K shown in Figure 2d; **e-h)** the under-focused Lorentz-TEM images showing magnetic-field-driven transitions from stripes (**d**) gradually to bubbles (**e-g**) and eventually ferromagnetic sate (**h**) with different external magnetic fields (H) applied along the *c*-axis at T=110 K. The scale bar is 500 nm.

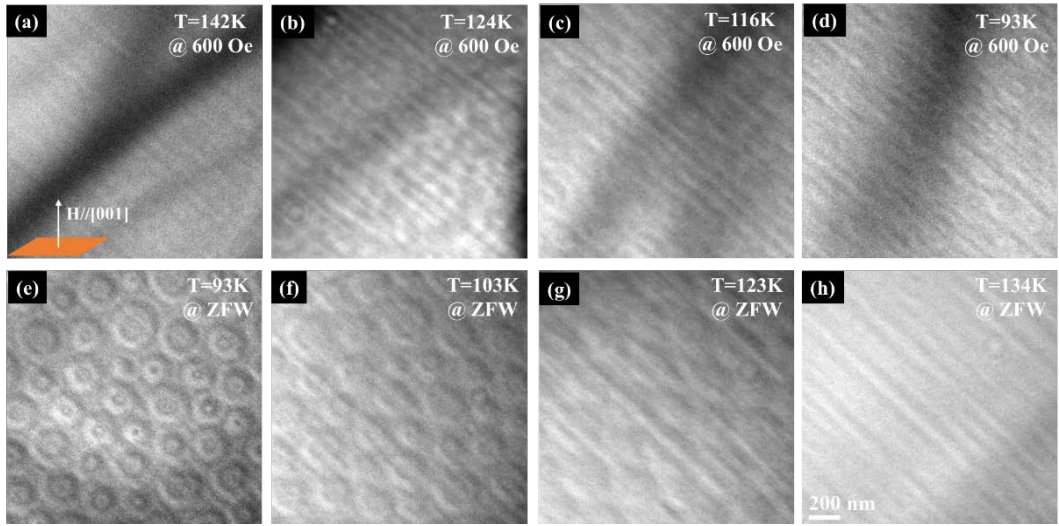

**Figure 3. a-d)** The under-focused Lorentz-TEM images of FGT when the sample temperature was lowered from 300 K to 93 K in a 600 Oe field-cooling (FC) process. The high-density skyrmion lattices is confirmed at temperatures below $T_C$ ~150 K after the FC procedure; **e-h)** The under-focused Lorentz-TEM images showing the bubble lattices evolution during a zero-field warming (ZFW) process. The scale bar is 200 nm.

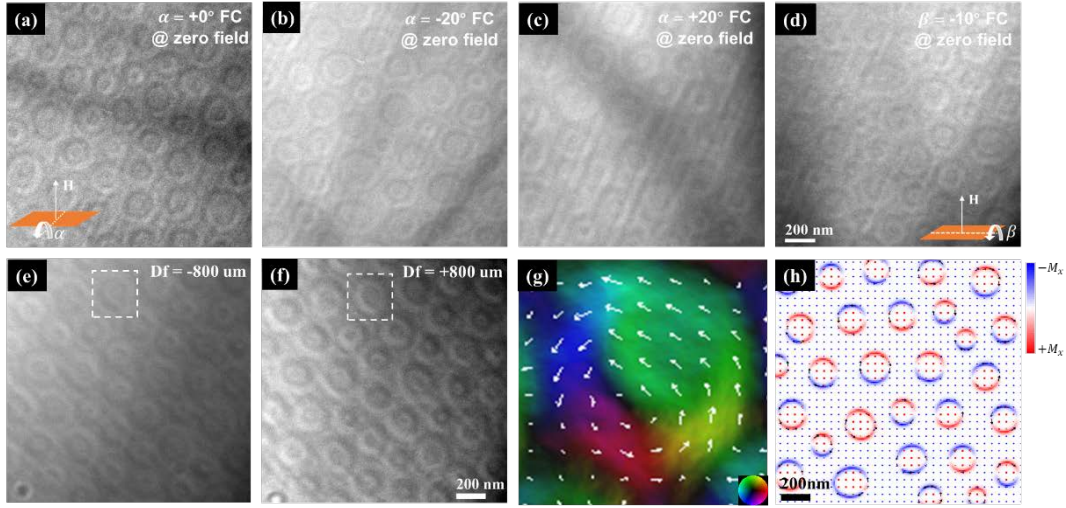

**Figure 4. a-d)** Under-focused Lorentz-TEM image of skyrmion bubbles at 93 K after FC manipulation with a field (600 Oe) applied with rotation angles. (α; as shown schematically in the inset of a) α = -20°, b) α = 0°, c) α = +20° and d) β = -10°, respectively). **e, f)** Under-focused and over-focused Lorentz-TEM images of the skyrmion bubbles taken at 93 K and in zero field; **g)** An enlarged in-plane magnetization distribution map obtained by TIE analysis for a selected skyrmion bubble indicated by white dotted box in (e) and (f). The white arrows represent magnetization direction at each point and the color wheel is in the right corner; **h)** Theoretical simulation of skyrmion lattices at an applied magnetic field with 600 Oe for α = 0°. In-plane magnetization distribution is represented by blue (−$m_x$) and red (+$m_x$) regions. The scale bar is 200nm.